\documentclass[runningheads]{llncs1}
\let\llncssubparagraph\subparagraph
\let\subparagraph\paragraph
\usepackage[compact]{titlesec}
\let\subparagraph\llncssubparagraph
\usepackage{graphicx}
\usepackage{caption}
\usepackage{subcaption}
\usepackage{algorithm}
\usepackage{algorithmic}
\usepackage{multirow}
\usepackage{multicol}
\usepackage{makecell}
\usepackage{array}
\usepackage{longtable}
\usepackage{booktabs}
\usepackage{cite}
\usepackage{amssymb}
\usepackage{amsmath}
\usepackage{xcolor}
\usepackage{setspace}
\usepackage{setspace}
\usepackage[colorlinks,bookmarksopen,bookmarksnumbered,citecolor=green, linkcolor=red, urlcolor=blue]{hyperref}

\begin{document}
\title{A Joint Approach to Local Updating and Gradient Compression for Efficient Asynchronous Federated Learning}


\author{Jiajun Song\inst{1} \and
Jiajun Luo\inst{2} \and
Rongwei Lu\inst{1} \and
Shuzhao Xie\inst{1} \and \\
Bin Chen\inst{3} \and
Zhi Wang\inst{1}\thanks{Corresponding author}
}
%
%
\institute{Shenzhen International Graduate School, Tsinghua University; \\ Tsinghua-Berkeley Shenzhen Institute, Tsinghua University \and
Department of Computer Science and Engineering,\\ Southern University of Science and Technology
\and
School of Computer Science and Technology, \\ Harbin Institute of Technology, Shenzhen\\
\email{sjj23@mails.tsinghua.edu.cn}, \email{12012023@mail.sustech.edu.cn},
\email{lrw21@mails.tsinghua.edu.cn},
\email{xiesz99@gmail.com},
\email{chenbin2021@hit.edu.cn},
\email{wangzhi@sz.tsinghua.edu.cn}
}
\begingroup
\let\clearpage\relax
\maketitle             
\begin{abstract}
Asynchronous Federated Learning (AFL) confronts inherent challenges arising from the heterogeneity of devices (e.g., their computation capacities) and low-bandwidth environments, both potentially causing \emph{stale} model updates (e.g., local gradients) for global aggregation. Traditional approaches mitigating the staleness of updates typically focus on either adjusting the local updating or gradient compression, but not both. Recognizing this gap, we introduce a novel approach that synergizes local updating with gradient compression. Our research begins by examining the interplay between local updating frequency and gradient compression rate, and their collective impact on convergence speed. The theoretical upper bound shows that the local updating frequency and gradient compression rate of each device are jointly determined by its computing power, communication capabilities and other factors. Building on this foundation, we propose an AFL framework called FedLuck that adaptively optimizes both local update frequency and gradient compression rates. Experiments on image classification and speech recognization show that FedLuck reduces communication consumption by $56$\% and training time by $55$\% on average, achieving competitive performance in heterogeneous and low-bandwidth scenarios compared to the baselines.

\end{abstract}
\section{Introduction}\label{introduction}

Asynchronous Federated Learning (AFL) \cite{asynchronousopt, fedperidic, fedbuff} is a machine learning method where devices train a model together without synchronizing \cite{fedopt}. However, this approach struggles with the varying local training speeds and capacities of devices, as well as low-bandwidth networks, leading to a problem known as \emph{model staleness} \cite{relatedwork-stale1, fedat, onlineafl,relatedwork-stale2,taylor}. The features of the asynchronous mechanism facilitate the emergence of model staleness due to the heterogeneous update speeds among different devices and communication time delays in low-bandwidth settings. The heterogeneity and limited bandwidth can cause certain devices to fall behind, leading to outdated local gradients that degrade the global model's performance. Solving model staleness is crucial for AFL deployment in real-life scenarios. 

Several strategies have been explored to mitigate this issue.  Some employ fusion weight methods \cite{asynchronousopt, asynwireless, stafl, fedperidic} for adaptive aggregation during global updates based on staleness levels.  Mathematical techniques like Taylor expansion and Hessian matrix approximation also mitigate staleness impact \cite{staleness-aware}. Besides, clustering devices \cite{safa, fedmds} by model staleness levels is also proven to be effective. Some studies \cite{liiclr, fedsa} focus on adjusting local updating frequency to accelerate convergence, but they overlook communication challenges, particularly in low-bandwidth scenarios where communication is often the bottleneck \cite{time}. To further reduce communication consumption, gradient compression algorithms \cite{dgc,topk,gaia,qsgd,terngrad,signsgd} are used to expedite transmission. However, many employ fixed compression rates without fully exploring the correlation between compression rates with other factors. Recent research has delved into this relationship, considering factors like device data volumes and network conditions \cite{dc2,dagc,optimalrate}. This has spurred the development of adaptive gradient compression algorithms, dynamically adjusting compression rates based on these factors.

Existing works on adjusting local updating frequency and compression rate \cite{fedsa, liiclr, fedperidic} have been proven effective. However, they do not adequately consider the joint optimization of both local updating and communication. Although beneficial in isolation, the lack of simultaneous consideration of both components may lead to inefficient solutions as these approaches overlook potential interactions between local updating, communication, and convergence speed. For instance, the gradient obtained through more local updates often requires a bigger compression rate to preserve more information. Besides, in either LANs or WANs, transferring a complete model takes $10-100 \times$ times longer than local updating time. Therefore,  considering local updating or gradient compression alone often results in low convergence speed. We need to consider both mechanisms to improve the convergence speed. Though  some synchronous FL methods \cite{ffl, fedlamp} have explored the joint optimization of local updating and gradient compression, AFL remains unexplored. Simply migrating synchronous optimization algorithms to AFL poses challenges such as inadequate low-speed device local training.

Our primary objective is to improve convergence speed through joint optimization of local updating and communication. To achieve this, we adjust two key parameters: the local updating frequency, which affects the time of local training, and the gradient compression rate, which influences communication time. However, this approach presents a dilemma. On one hand, high local updating frequency and compression rate lead to significant model staleness, hindering the convergence of the global model; On the other hand, low values for these parameters cause significant compression errors, impeding the accuracy of the global model. Our experimental motivation reveals that with appropriate settings, convergence occurs $11 \times$ faster than with the most inappropriate settings in average. 

Therefore, a critical question emerges: \textbf{Given a specific AFL system, how can we jointly and adaptively decide the local updating frequency and the compression rate to converge efficiently?} In this paper, we undertake a theoretical analysis to reveal the complex interplay between the two parameters and their impact on convergence speed. Our analysis leads to the derivation of an upper bound for convergence speed based on the two parameters.  Based on this upper bound, we introduce an optimization problem tailored to determine these values for the two parameters. Building upon these insights, we introduce \textbf{FedLuck}, an AFL framework designed to enhance convergence efficiency by adjusting both local updating frequency and gradient compression rate.  The main contributions are summarized as follows:
\vspace{-2.8pt}
\begin{itemize}
\item We propose FedLuck, a novel AFL framework to enhance convergence efficiency by jointly and adaptively deciding the local updating frequency and gradient compression rate.
    
\item We conduct a theoretical analysis of the convergence speed and derive an upper bound with respect to the local updating frequency and compression rate.  We find a \textbf{key convergence factor $\phi$}, which exerts an impact of the largest magnitude on the convergence speed. We implemented FedLuck which jointly and adaptively decides the local update frequency and the compression rate for each device by minimizing $\phi$.
    
\item We extensively evaluated the performance of FedLuck through simulations and testbed experiments. The evaluation results demonstrate that FedLuck achieves significant reductions in communication consumption by $56$\% and completion time by  $55$\% on average while maintaining competitive performance compared to the baselines.
\end{itemize}
\section{Preliminaries and Problem Formulation}\label{preliminaries}

\subsection{Federated Learning}\label{subsec:2.1FL}
This paper focuses on a parameter server (PS) architecture, comprising a centralized PS for global aggregation and a set of $N$ distributed local devices represented as $\mathcal{N}$. The PS maintains a global model $\mathbf{w} \in \mathbb{R}^d$. Each device, represented by $i \in \mathcal{N}$, possesses a model weight $\mathbf{w}_{i}$ and a local dataset $D_i$. Within this dataset, there exist $|D_i|$ data samples, expressed as $\xi_i = [\xi_{i,1}, \xi_{i,2}, \cdots, \xi_{i,|D_i|}]$, which are utilized for local training. We define the loss function for each data sample $\xi_{i,j}$ ($j \in [1,|D_i|]$), as $f(\mathbf{w}_i, \xi_{i,j})$,  and denote the local loss function of device $i$ as:

\begin{equation}
    F_i(\mathbf{w}_i) := \frac{1}{|D_i|} \sum_{j = 1}^{|D_i|} f(\mathbf{w}_i, \xi_{i,j}), 
    \label{local loss function}
\end{equation}
where the global loss function can be formulated as $F(\mathbf{w}) := \frac{1}{N}\sum_{i \in \mathcal{N}}  F_i(\mathbf{w}).$

The target of this system is to train a global model $\mathbf{w}^{*}$ that minimizes the global loss function:
\begin{equation}
    \mathbf{w}^{*} := \arg \mathop{\min}_{\mathbf{w} \in \mathbb{R}^d} F(\mathbf{w}).
\end{equation}

In FL, the whole training process can be divided into model broadcasting, local training, and model aggregation. 
This iterative update process continues for multiple rounds, allowing the global model to be refined and improved over time.

\subsection{AFL with Periodic Aggregation}\label{subsec:2.2AFL}

We specifically target AFL with periodic aggregation. The server periodically aggregates the received gradients from devices that have completed their computations, while allowing other devices to continue their local training uninterrupted. The interval between two global rounds remains constant. Once a device completes its local training, it transmits its local gradient to the server. 


Specifically, the entire training process comprises a total of $T$ global rounds. Within each global round, every local device $i$ undertakes $k_i$ local iterations. At each local iteration $j$, ranging from 0 to $k_i - 1$, local device $i$ updates its local model following the rule:
\begin{equation}
    \mathbf{w}_i^{t, j+1} = \mathbf{w}_i^{t, j} - \eta_l \nabla F_i(\mathbf{w}_i^{t,j}, \xi_i^{t,j}),
\end{equation}
where $\eta_l$ is the learning rate of local device, and $\mathbf{w}_i^{t, j}$ is the model of $j$-th local iteration of device $i$ training with global model $\mathbf{w}^t$.

When a local device $i$ has finished its $k_i$ local updates to train the global model $\mathbf{w}^t$, it computes the overall gradient $\mathbf{g}_i^t$ in local training, that is:
\begin{equation}
    \mathbf{g}_i^t = \mathbf{w}_i^{t, 0} - \mathbf{w}_i^{t, k_i}.
\end{equation}

Note that when a local device $i$ receives the $t$-th global model $\mathbf{w}^t$, it will be initialized with $\mathbf{w}_i^{t, 0} = \mathbf{w}^t$. Then, the local device will upload the compressed gradient to the PS. In this paper, we employ $top_{k}$ sparsification algorithm \cite{topk} as the compressor and other compressors can also be employed. 


The compression rate of the $top_k$ compression operator is defined as $\delta = \frac{k}{d}$, where $k$ represents the number of parameters retained after compression, and $d$ denotes the total number of original parameters. After compressed by $top_{k}$ compressor, the device $i$ will send the compressed gradient $\tilde{\mathbf{g}}_i^t = C_{\delta_i} (\mathbf{g}_i^t)$ to the parameter server. We then use $C_{\delta_i}$ to express the $top_{k}$ compressor with a compression rate of $\delta_i$. The amount of data transmitted is proportional to the compression rate, so the time for transmitting a compressed gradient is also proportional to the compression rate.

Considering the diversity in computational capabilities and communication capabilities among devices, the total time for training and communication of device $i$ is:
\begin{equation}
\label{equ:device time}
    d_i = k_i\alpha_i + \delta_i\beta_i,
\end{equation}
where the local updating frequency is $k_i$ and the compression rate is $\delta_i$. We define $C_{\delta_i}$ as the compressor with compression rate $\delta_i$. Let $\alpha_i$ denote the computation time required for one local update on device $i$, and $\beta_i$ represent the communication time for transmitting a full-size gradient on device $i$. Given that the download bandwidth is typically higher than the upload bandwidth \cite{uploadtime1,fedlamp}, our attention is primarily directed toward the communication time involved in transmitting the models from devices to the PS during the model exchange process.
 
During local training and communication, the server continuously receives gradients from local devices. We define $\mathbf{S}^t$ as the set of local devices to which the server has received gradients in the $t$-th global round. The PS then aggregates the received local gradients from $\mathbf{S}^t$ and updates the global model:

\begin{equation}
    \label{equ:global update}
    \mathbf{w}^{t + 1} = \mathbf{w}^{t} - \frac{\eta_g}{|\mathbf{S}^t|} \sum_{i \in \mathbf{S}^t}  \tilde{\mathbf{g}}_i^{t - \tau_i^t},
\end{equation}
where $\eta_g$ is the global learning rate. Due to the asynchronous nature, the gradient may be stale. That is, the gradient of device $\tilde{\mathbf{g}}_i^{t - \tau_i^t}$ is generated by device $i$ to train the global model $\mathbf{w}^{t - \tau_i^t}$. When the PS receives $\tilde{\mathbf{g}}_i^{t - \tau_i^t}$, it is executing the $t$-th round of aggregation. Building on the preceding analysis, the staleness of device $i$ to train the global model $\mathbf{w}^{t - \tau_i^t}$ is $\tau_i^t$ that is defined as:
\begin{equation}
\label{equ:staleness}
    \tau_i^t = t - \max_{t^{'} < t} \{t^{'} | i \in \mathbf{S}^{t^{'}} \}.
\end{equation}

According to the characteristics of AFL with periodic aggregation, the time interval between every two global rounds is fixed. Therefore, to employ the staleness $\tau_i^t$ in the theoretical analysis, $\tau_i^t$ can be also calculated as $\lceil \frac{d_i}{\tilde{T}}\rceil$, where $\tilde{T}$ is the duration of one round. This equivalence holds when the values of local updating frequency and compression rate are constant for each device.
\section{The Joint Approach to Local Updating and Gradient Compression}\label{FedLuck}

In this section, we first explore the impacts of both local updating frequency and compression rate, providing two motivation experiments to illustrate the dilemma discussed in Sec. \ref{introduction}. Then, we derive the upper bound of the convergence rate through theoretical analysis, leading to the proposal of FedLuck, an AFL framework that jointly and adaptively decide the local updating frequency and the compression rate for enhanced training efficiency.

\subsection{Illustrative Examples of Motivation}\label{motivation example}

As mentioned in Sec. \ref{introduction}, model staleness in AFL is affected by the time each device takes in one round, comprising local updating and communication time. We regulate local updating time with the local updating frequency and communication time with the compression rate. Our exploration reveals optimal values for both parameters, as elaborated below.

\begin{figure}[htbp]
\setlength{\belowcaptionskip}{-3.5mm}

  \centering
  \begin{subfigure}[b]{0.4\textwidth}
    \includegraphics[width=\textwidth]{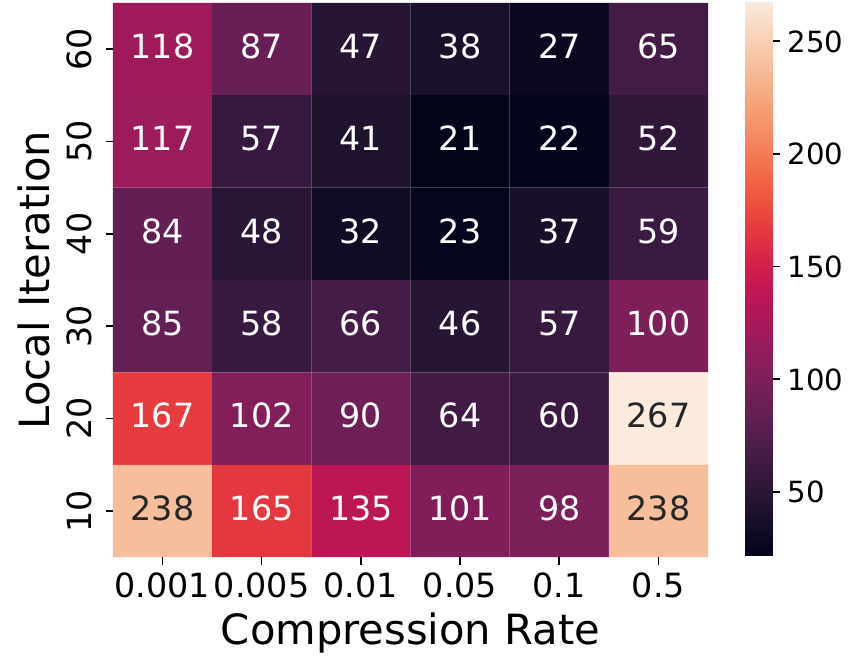}
    \caption{CNN@FMNIST}
    \label{fig:heatmap_cnn}
  \end{subfigure}
  \hspace{0.2cm} 
  \begin{subfigure}[b]{0.4\textwidth}
    \includegraphics[width=\textwidth]{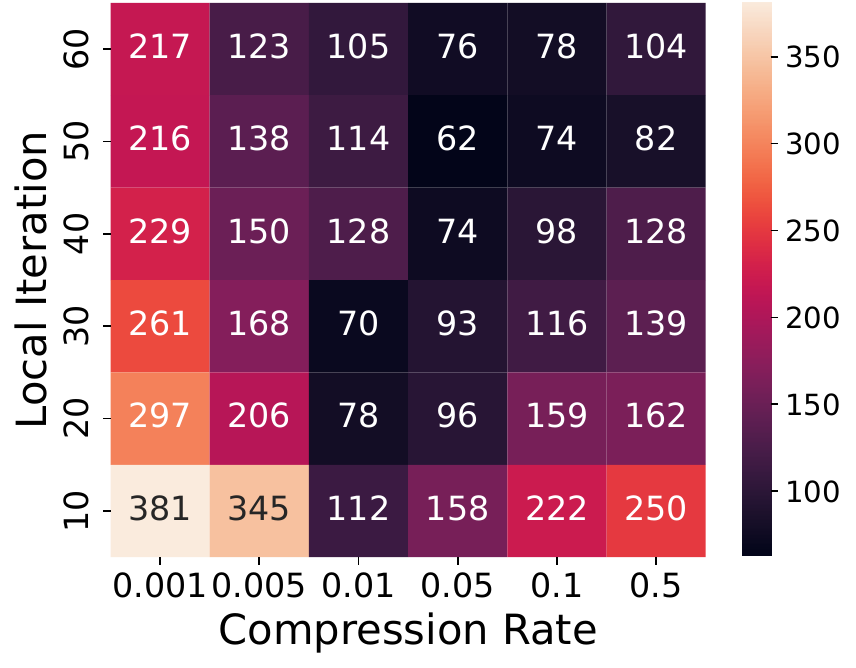}
    \caption{LSTM@SC}
    \label{fig:heatmap_lstm}
  \end{subfigure}
  \caption{The minimum global rounds required to reach the target accuracy for the global model under different compression rates and local updating frequencies.}
  \label{fig:heatmap}
\end{figure}

We conducted both image classification (training CNN on Fashion-MNIST dataset \cite{fmnist}, abbreviated as FMNIST) and speech recognition (training LSTM on Speech Commands dataset \cite{sc}, abbreviated as SC) experiments, which is sufficient to validate our motivation. We used 5 devices and a PS for both experiments. We set the local updating frequency to integers multiple of ten between 10 and 60, and the compression rate to 0.001, 0.005, along with 10 and 100 times these values. The detailed results are depicted in Fig. \ref{fig:heatmap} which shows the minimum number of global rounds required for the global model to achieve an accuracy of 0.85 and 0.5. Specifically, deviating from the experimentally determined efficient settings either by increasing or decreasing can lead to a convergence process that is up to 3 times slower than the rate achieved at the fastest observed settings.

\subsection{Our Proposed AFL Framework: FedLuck}\label{Theoretical}
\textbf{Theoretical Analysis.} Once we have confirmed our motivation, we must address the following problem: Given a specific AFL system, how can we jointly and adaptively decide the local updating frequency and the compression rate to converge efficiently?

We delve into the examination of the convergence bound of the global loss function over $T$ global iterations. To commence, we outline the assumptions imposed on the training model, which are commonly embraced in pioneering federated learning works  \cite{sharper} and similar to  \cite{fedbuff}.

\textbf{Assumption 1.} (Bounded local variance). \textit{There exists a constant $\sigma$, such that the variance of each local estimator is bounded by: }
\begin{equation}
    \mathbb{E}_{\xi \sim D_i} \big[ ||\nabla F_i(\mathbf{w}, \xi) - \nabla F_i(\mathbf{w})||\big] \le \sigma, \forall i \in \mathcal{N}, \forall \mathbf{w} \in \mathbb{R}^d.
\end{equation}

\textbf{Assumption 2.} (Bounded function heterogeneity). \textit{There exist $N$ constants $\zeta_i^2 \ge 0, i \in \{1,2,\ldots,N\}$, such that the variance of the model gradients is bounded by:}
\begin{equation}
    ||\nabla F_i(\mathbf{w}) - \nabla F(\mathbf{w})||^2 \le \zeta_i^2, \forall \mathbf{w} \in \mathbb{R}^d,
    \label{ass:noniid}
\end{equation}
and we define $\zeta^2 := \frac{1}{N}\sum_{i \in \mathcal{N}} \zeta_i^2$.

\textbf{Assumption 3.} (L-smooth). \textit{The loss functions $F$ and $F_i$ are L-smooth with a constant $L \ge 0$ such that:}
\begin{equation}
    ||\nabla F_i(\mathbf{y}) - \nabla F_i(\mathbf{x})|| \le L||\mathbf{y} - \mathbf{x}||, \forall \mathbf{x}, \mathbf{y} \in \mathbb{R}^d.
\end{equation}

\textbf{Assumption 4.} (Bounded gradient). \textit{There exists a constant $G \ge 0$ such that the norm of local gradient is bounded by:}
\begin{equation}
    ||\nabla F_i(\mathbf{w})||^2 \le G^2, \forall \mathbf{w} \in \mathbb{R}^d.
\end{equation}

These assumptions are widely adopted and considered standard in the context of non-convex optimization problems. We show the detailed theoretical analysis in \href{https://github.com/Aegsteh-T/FedLuck_Euro_Par}{https://github.com/Aegsteh-T/FedLuck\_Euro\_Par}.

\textbf{Theorem 1.} If all local models are initialized by the same weight $\mathbf{w}^{0}$. Assuming the global learning rate and local learning rate $\eta_g, \eta_l \le \frac{1}{L}$ and $Q_l = 1 - 6(1 - \eta_l k)^2 \ge 0$, the mean square of gradient after $T$ global iterations is bounded as follows:
\begin{equation}
\label{theorem1}
\begin{split}
\frac{1}{T}&\sum_{t = 0}^{T - 1}||\nabla F(\mathbf{w}^{t})||^2 
\le \frac{2[F(\mathbf{w}^{0}) - F(\mathbf{w}^{*})]}{\eta_g Q_lT}
\\
&+ \frac{\eta_l^2k^2([(1 - \delta)B_1 + B_2]}{Q_l} 
+ \frac{\eta_l^2k^4[2\tau_{max}^2(2 - \delta) + 1]B_1}{Q_l},
\end{split}
\end{equation}
where $B_1 = 18(\sigma^2 + \zeta^2 + G^2)$ and $B_2 = 6(\sigma^2 +\zeta^2)$ represent the bias terms in the system. $\tau_{max} = \mathop{\max}_{i \in \mathcal{N}} \; \tau_i$ is the max model staleness of all devices. 

\textbf{Corollary 1.} Choosing global learning rate $\eta_g = \mathcal{O}(\sqrt{\frac{k}{T}})$ and local learning rate $\eta_l = \mathcal{O}(T^{-1/4}k^{-5/2}\delta^{-1/4})$, we have the convergence rate:
\vspace{-0.5em}
\begin{equation}
\label{corollary1} 
\begin{split}
\frac{1}{T}\sum_{t = 0}^{T - 1}&||\nabla F(\mathbf{w}^{t})||^2 
\le
\mathcal{O}(\frac{F^{*}}{\sqrt{kT}}) 
+ \mathcal{O}(\frac{(1 - \delta)B_1+B_2}{k^3\sqrt{T\delta}})
\\
&+ \mathcal{O}(\frac{(\tau_{max}^2(2 - \delta) + 1)B_1}{k\sqrt{T\delta}}).
\end{split}
\end{equation}

Given that the third term exerts the highest order of influence on the convergence speed, our focus is directed toward this term to discuss strategies for achieving efficient convergence. We name the part related to local updating frequency and compression rate in the third term as \textbf{key convergence factor}. To get the fastest convergence, we should minimize the key convergence factor. According to the definition of staleness, we expand $\tau_{max} = \frac{k\alpha + \delta \beta}{\tilde{T}}$, where $\alpha = \alpha_m, \beta = \beta_m, k = k_m, \delta = \delta_m,  m = \arg \mathop{\max}_{i} \; \tau_{i}$. Then we get the key convergence factor as:
\begin{equation}
    \phi(k, \delta) = \frac{(k\alpha + \delta\beta)^2(2 - \delta) + \tilde{T}^2}{\tilde{T}^2 k\sqrt{\delta}}.
    \label{dominant term}
\end{equation}

To extend the key convergence factor to all devices, we aim to find the optimal solutions of local updating frequency $k_i$ and compression rate $\delta_i$ for each device $i$ to achieve fast convergence. This optimization problem is formulated to minimize the key convergence factor $\phi(k_i, \delta_i)$ while considering constraints on $k_i$ and $\delta_i$ within specific ranges. By solving this heuristic optimization problem, we obtain adaptive values for $k_i$ and $\delta_i$ for each device, enabling efficient convergence in the training process. The optimization problem is formulated as follows:

\begin{equation}
\begin{split}
\min_{k_i, \delta_i} &\;\; \phi(k_i,\delta_i)
\\
s.t. &\;  k_i \in [k_{min}, k_{max}], \quad  \delta_i \in [\delta_{min}, \delta_{max}],
\end{split}
\label{optimization problem}
\end{equation}
where $k_{min}$ and $k_{max}$ are the minimal local updating frequency and maximal local updating frequency, respectively. $\delta_{min}$ and $\delta_{max}$ are the similar meanings for compression rates. 
\vspace{-2em}
\begin{algorithm}\footnotesize
\caption{\footnotesize{FedLuck}}
\setstretch{0.85}
\label{alg:fedluck}
\begin{multicols}{2}
\begin{algorithmic}[1]
\STATE \textbf{Server:}
    \STATE Receive $\alpha_i$ from devices
    \STATE Test $\beta_i$ with devices
    \STATE Get the optimal $k_i$ and $\delta_i$ according to Eq. \ref{optimization problem}
    \STATE Send ($k_i$, $\delta_i$) to device $i$
    \STATE Broadcast $\mathbf{w}^0$ to all devices and start them
    \FOR{$t = 0,1, \cdots, T - 1$:}
        \STATE Continuously receive $\tilde{\mathbf{g}}_i^{t - \tau_i^t}$ from local devices set $\mathbf{S}^t$
        \STATE  $\mathbf{w}^{t + 1} \gets \mathbf{w}^{t} - \frac{\eta_g}{|\mathbf{S}^t|} \sum_{i \in \mathbf{S}^t}  \tilde{\mathbf{g}}_i^{t - \tau_i^t}$
        \FOR{$i \in \mathbf{S}^t$}
        \STATE Send new global model $\mathbf{w}^{t + 1}$ to $i$
        \ENDFOR
    \ENDFOR
    \STATE Notify all devices to \textit{STOP}
    
    \STATE \textbf{Device:}
    \STATE Test $\alpha_i$ and send it to the server
    \STATE Test $\beta_i$ with the server 
    \STATE Receive ($k_i$, $\delta_i$) from server
    \WHILE{not \textit{STOP}}
    \STATE Receive $\mathbf{w}^t$ from server
    \STATE Set $\mathbf{w}_i^{t,0} \gets \mathbf{w}^t$
    \FOR{each local iteration $j \in {0,1,\cdots,k_i - 1}$:}
    \STATE $\mathbf{w}_i^{t,j+1} \gets \mathbf{w}_i^{t,j} - \eta_l \nabla F_i(\mathbf{w}_i^{t,j};\xi_i^{t,j})$
    \ENDFOR
    \STATE Compute gradient $\mathbf{g}_i^t \gets \mathbf{w}_i^{t,0} - \mathbf{w}_i^{t,k_i}$
    \STATE Compute compressed gradient $\tilde{\mathbf{g}}_i^t \gets C_{\delta_i}(\mathbf{g}_i^t)$
    \STATE Send $\tilde{\mathbf{g}}_i^t$ to server
    \ENDWHILE
\end{algorithmic}
\end{multicols}
\end{algorithm}
\vspace{-2em}
Based on the formula (Eq. \ref{optimization problem}), we can find the adaptive values $k_i$ and $\delta_i$ for each device $i$. We show FedLuck in Alg. \ref{alg:fedluck}. Devices must determine their local updating frequency and compression rate before training by conducting tests on local updating and communication time. Local training rounds are performed on each device to calculate the average local training time per iteration, denoted as $\alpha_i$. Devices also measure gradient transmission time by sending compressed gradients with varying timestamps and compression rates, allowing the server to compute the communication time required for a full gradient, denoted as $\beta_i$. The server then solves the optimization problem (Eq. \ref{optimization problem}) to find the values for $k_i$ and $\delta_i$, which are sent to each device. Devices apply these parameters for AFL. A comparison of our approach with other baselines is presented in the subsequent section.

\section{Experiments}\label{experiments}

In this section, we compare our proposed method, FedLuck, with four baselines. We outline the experimental settings, encompassing the tasks, baselines, and metrics. We also detail the hardware and hyper-parameter configurations used in our experiments. Finally, we present the results under empirical settings.

\subsection{Experiment Tasks}

\textbf{Tasks.} In our experiments, we evaluated the performance of our FedLuck framework using three diverse tasks: (1) a four-layer CNN model for FMNIST (Fashion-MNIST \cite{fmnist}) dataset for image classification, inspired by~\cite{federatedchallenge}, (2) VGG11s  \cite{vgg} on CIFAR-10\footnote{http://www.cs.toronto.edu/\textasciitilde kriz/cifar.html} for image classification. VGG11s' architecture is similar to \cite{stc}, and (3) an LSTM for SC (Speech Commands, \cite{sc}). The LSTM features two hidden layers with 128 units each. These models are tested across both visual and auditory domains to demonstrate the versatility of our approach.

\subsection{Baselines and Metrics}
\textbf{Baselines.} We selected four classical algorithms as baselines for performance comparison. (1) \textbf{FedPer:}  An AFL method with periodic aggregation, maintaining fixed and uniform local updating frequency and consistent compression rate across devices \cite{fedperidic}. (2) \textbf{FedBuff:} An AFL method that aggregates a fixed number of gradients in each global round \cite{fedbuff}. (3) \textbf{FedAsync:} FedAsync  \cite{asynchronousopt} is an AFL method that incorporates a regularization term to enhance performance when dealing with staleness. Moreover, FedAsync changes fusion weight with respect to model staleness. (4) \textbf{FedAvg + Topk:} Based on the FedAvg \cite{fedavg} algorithm, we have incorporated the Topk gradient compression technique to mitigate communication overhead  \cite{topk}.

\textbf{Metrics.} We assess and compare the performance of FedLuck with the other baselines using the following metrics. (1) \textbf{Test accuracy} is the ratio of correct predictions to the total test data points. (2) \textbf{Communication consumption} measures the data volume exchanged in federated learning, assessing communication efficiency.

\subsection{Simulation Experiments}
\textbf{Simulation Settings.} We conducted simulation experiments to evaluate the performance of FedLuck on a server running Ubuntu 18.04.6 LTS, equipped with an Intel(R) Xeon(R) Silver 4210 CPU (20 cores, 40 threads) and 4 Nvidia GeForce GTX 3090 GPUs with CUDA version 11.2. We implemented the model training for each device on PyTorch, with Python version 3.8.13 and Torch version 1.12.0. 
We consider a common scenario where devices communicate with the central PS through a low-bandwidth network. There are 10 devices and 1 PS in total. Our simulation focuses on a typical scenario where each device's bandwidth ranges from $0.25$ Mb/s to $2$ Mb/s \cite{fedavg}. In addition, to simulate the heterogeneity of computation, we assume that the computation time of a local iteration of a simulated device follows a uniform distribution. The minimum value of the uniform distribution is multiple times of the computation time of one local iteration on the Nvidia GeForce GTX 3090 24GB GPU. The maximum value of the uniform distribution is four times the minimum value.

We utilized the momentum-SGD optimizer across all three tasks, initializing the learning rates to 0.01 for the CNN model on FMNIST and VGG11s on CIFAR-10, and 0.016 for the LSTM model on SC, with a momentum of 0.9 for all models. A batch size of 64 was employed for the LSTM and CNN models, while a batch size of 32 was used for the VGG model, to efficiently process the training data. These settings were carefully selected to enhance the training process and ensure reliability during evaluation.

\begin{figure*}[!t]
\setlength{\belowcaptionskip}{-3.5mm}
    \centering
    \subfloat[CNN@FMNIST]{
        \includegraphics[width=0.3\textwidth]{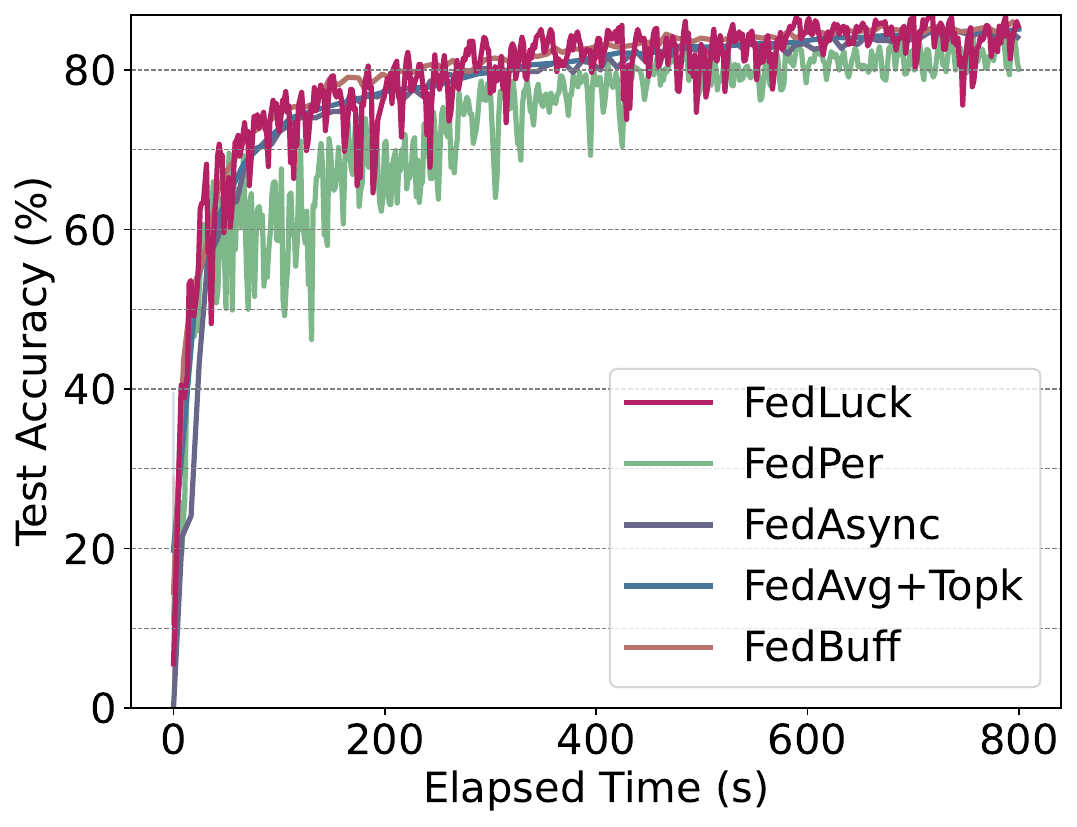}
        \label{fig:FMNIST_iid_time_acc}
    }
    \hfill
    \subfloat[VGG11s@CIFAR-10]{
        \includegraphics[width=0.3\textwidth]{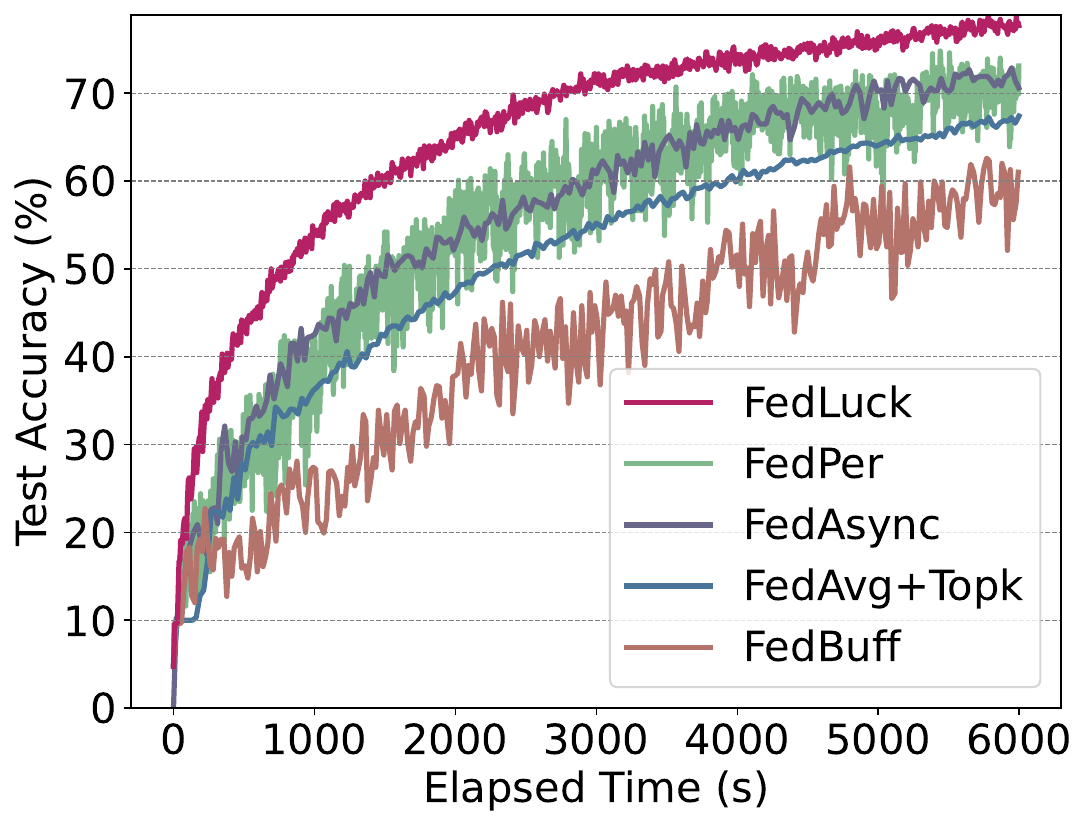}
        \label{fig:CIFAR10_iid_time_acc}
    }
    \hfill
    \subfloat[LSTM@SC]{
        \includegraphics[width=0.3\textwidth]{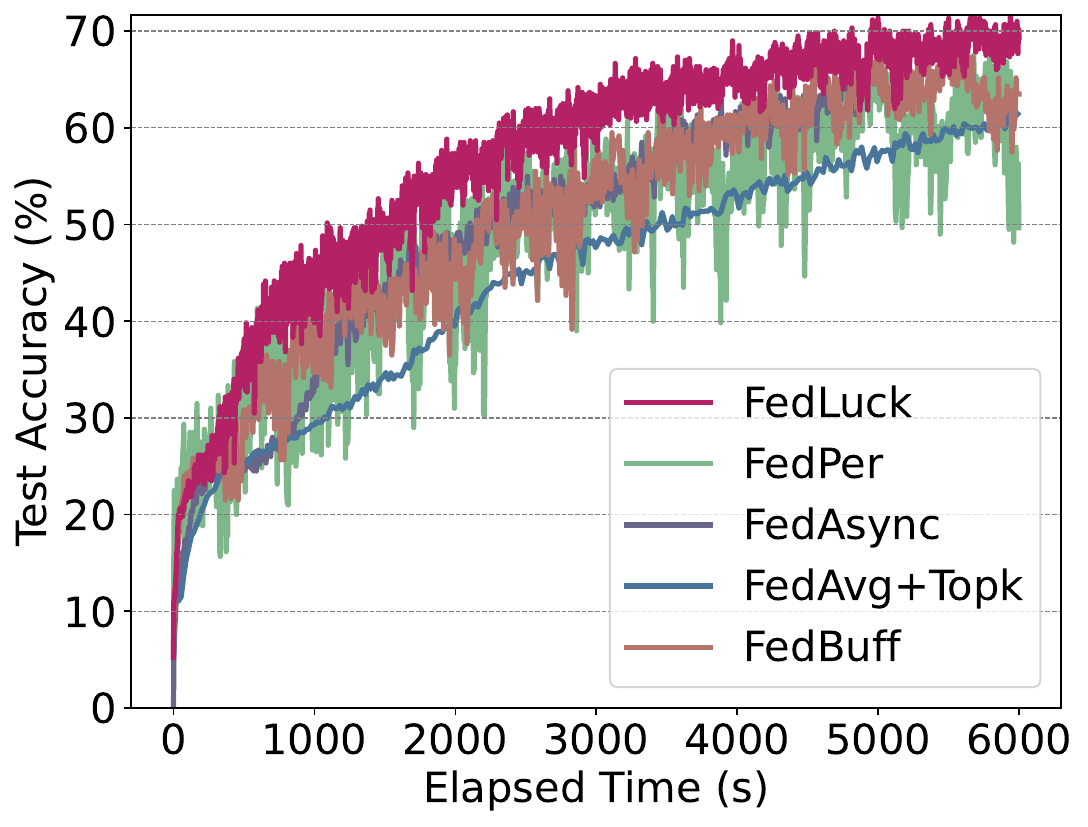}
        \label{fig:SC_iid_time_acc}
    }
    \caption{Test Accuracy and Elapsed Time comparison between FedLuck and four baselines on three tasks in IID setting. FedLuck reduces training time by  $55$\% in average compared with baselines.}
    \label{fig:iid_time_acc}
\end{figure*}

\begin{figure*}[!t]
\setlength{\belowcaptionskip}{-3.5mm}
    \centering
    \subfloat[CNN@FMNIST]{
        \includegraphics[width=0.3\textwidth]{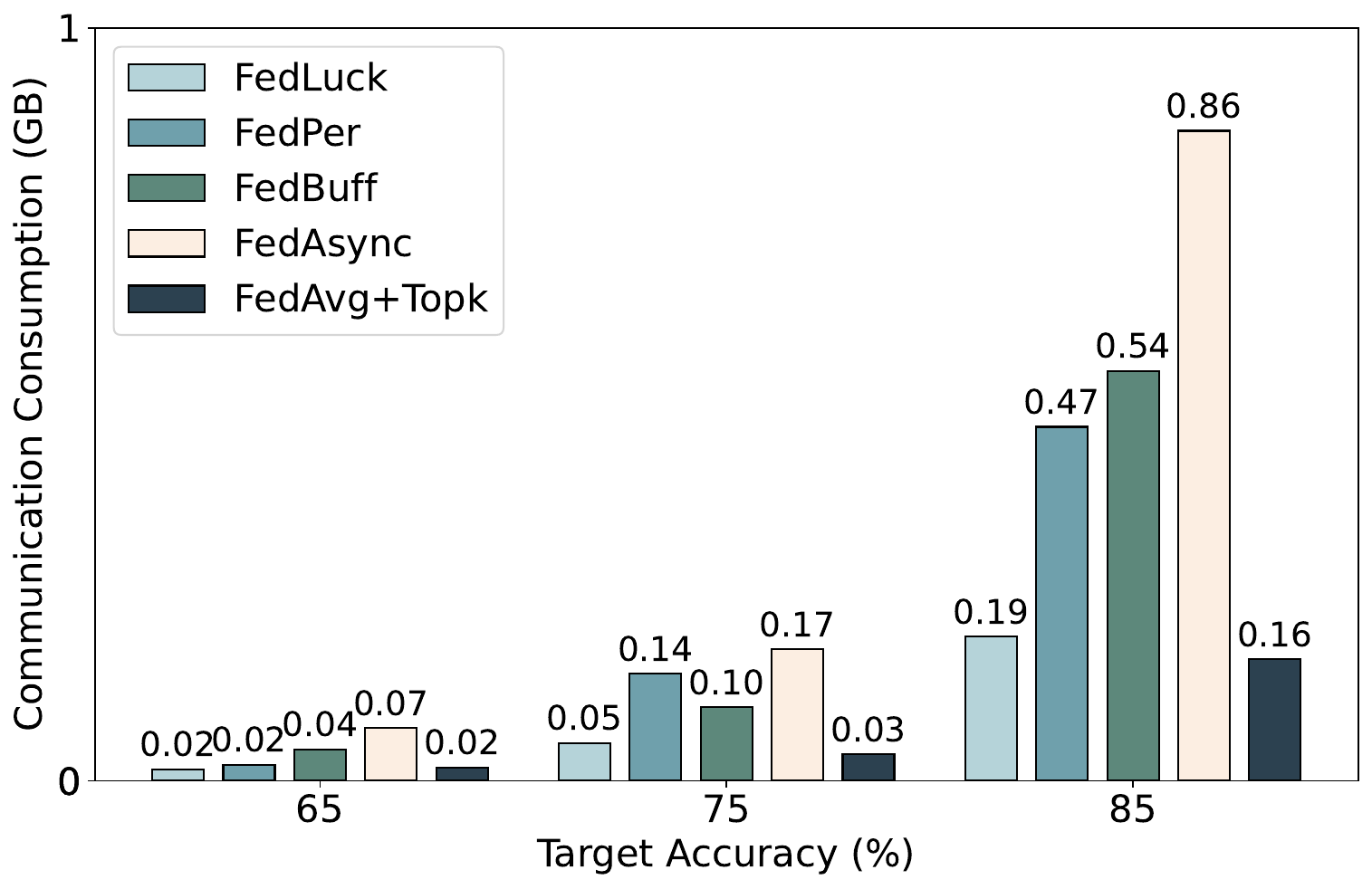}
        \label{fig:FMNIST_iid_acc_cost}
    }
    \hfill
    \subfloat[VGG11s@CIFAR-10]{
        \includegraphics[width=0.3\textwidth]{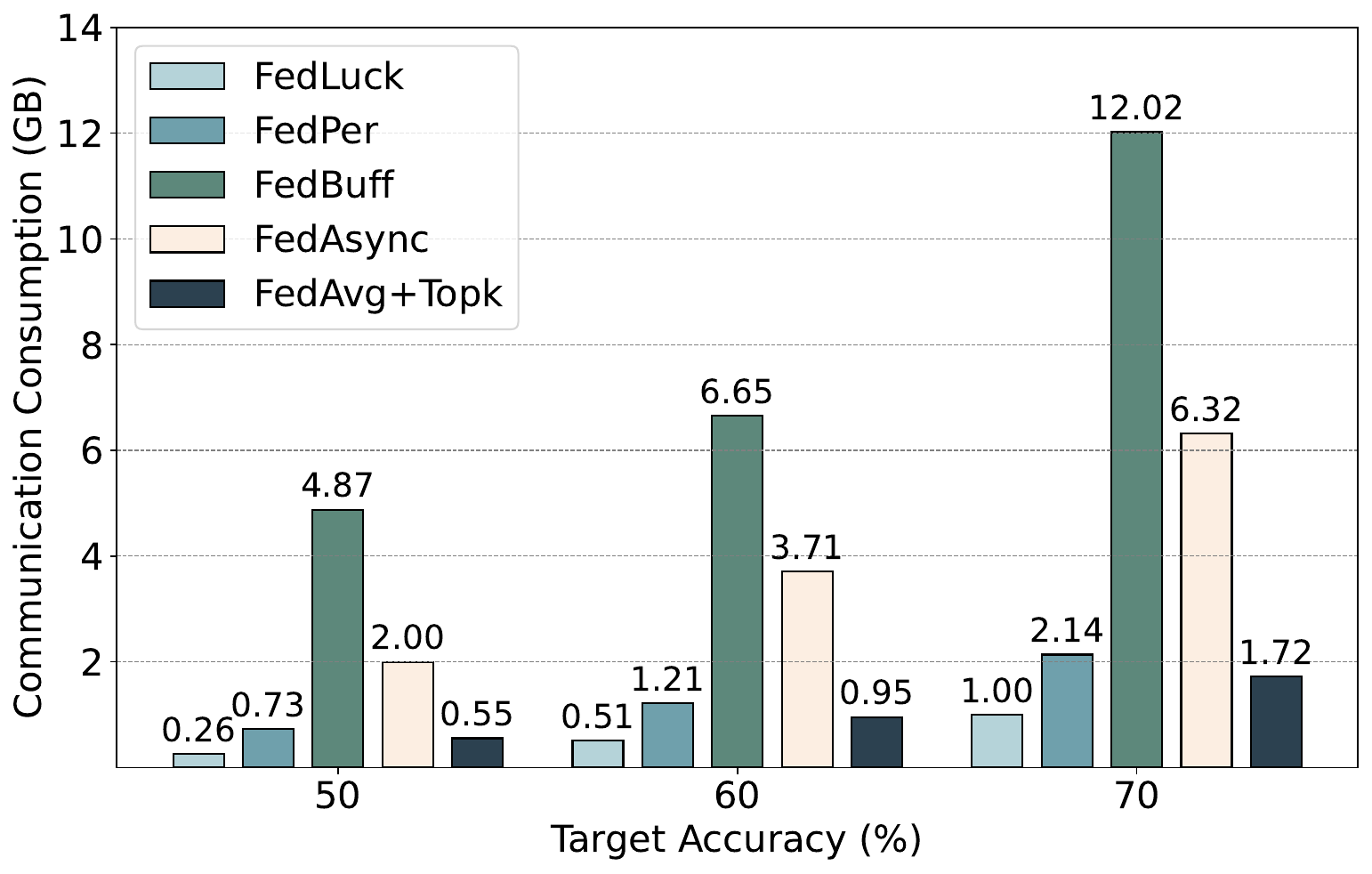}
        \label{fig:CIFAR10_iid_acc_cost}
    }
    \hfill
    \subfloat[LSTM@SC]{
        \includegraphics[width=0.3\textwidth]{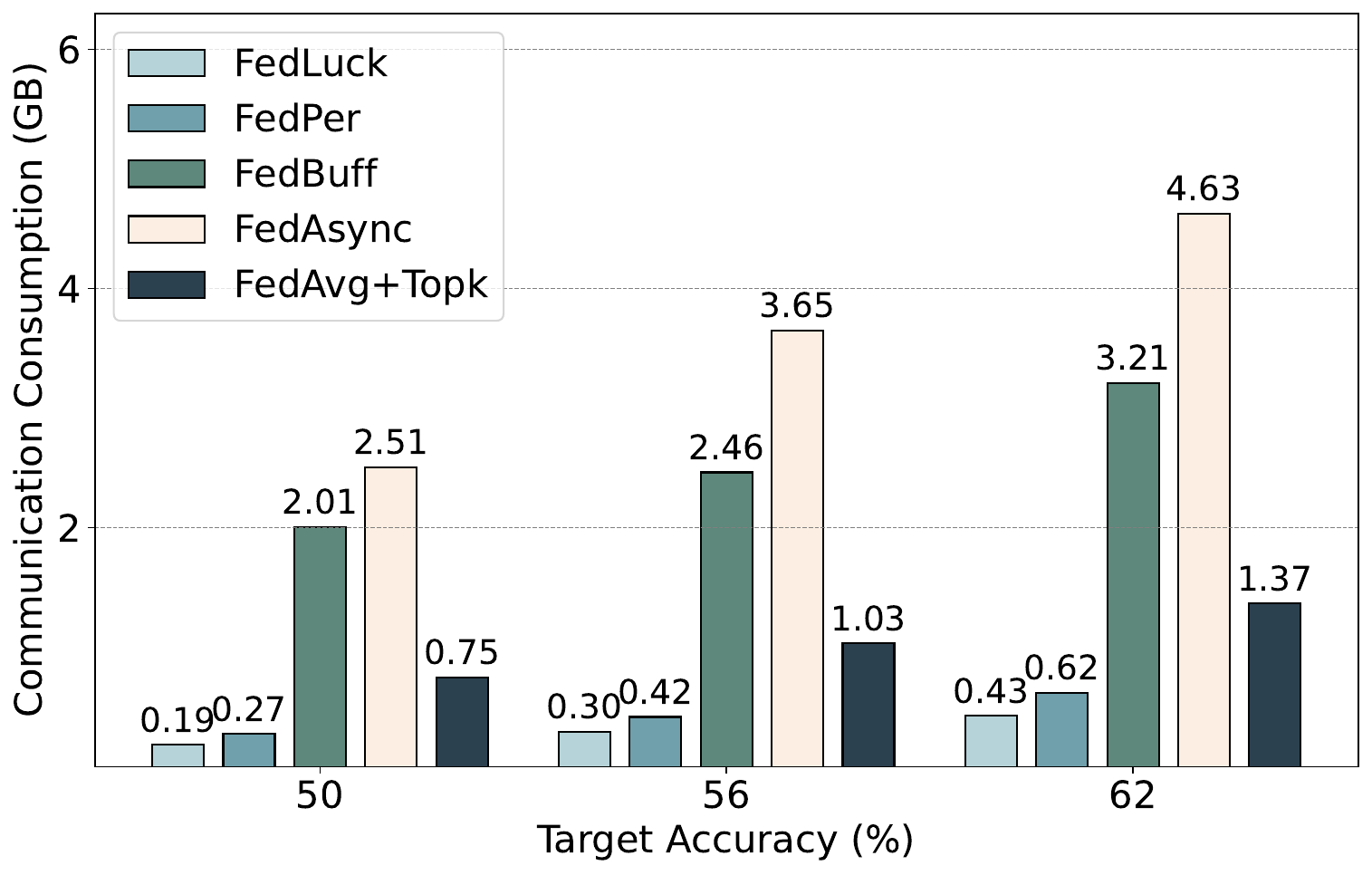}
        \label{fig:SC_iid_acc_cost}
    }
    \caption{Communication consumption between FedLuck and four baselines at target accuracy on three tasks in IID setting. FedLuck reduces communication consumption by $56$\% in average compared with baselines.}
    \label{fig:iid_com_acc}
\end{figure*}

\textbf{Simulation Results.} We evaluated the efficiency of FedLuck in terms of convergence speed and communication consumption. The horizontal axis represents the training time, while the vertical axis represents the test accuracy of the global model on the test dataset. This result highlights the effectiveness of jointly optimizing local updating frequencies and gradient compression rates for heterogeneous devices. As shown in Fig. \ref{fig:SC_iid_time_acc}, FedLuck takes 2517.61s to reach 62\% accuracy for LSTM on SC, while FedPer, FedBuff, FedAsync, FedAvg+Topk requires 3438.00s, 3452.12s, 3887.05s, 6136.02s, respectively. These timings translate to training speedups of 1.36$\times$, 1.37$\times$, 1.54$\times$, 2.44$\times$ for FedLuck compared to FedPer, FedBuff, FedAsync, FedAvg+Topk.

To illustrate the communication efficiency of FedLuck, we present the network communication consumption for different algorithms as they attain varying target accuracies, detailed in Fig. \ref{fig:iid_com_acc}. FedLuck demonstrates the most efficient network communication consumption in 6 out of 9 test cases. Even in the remaining three cases where it's not the best, FedLuck's performance is notably close to the best result. In Fig. \ref{fig:CIFAR10_iid_acc_cost} for VGG11s on CIFAR-10, when achieving the 70\% target accuracy, FedLuck, FedPer, FedBuff, FedAsync and FedAvg+Topk are 1.00GB, 2.14GB, 12.02GB, 6.32GB and 1.72GB, respectively. Concretely, FedLuck can save network communication consumption by 53\%, 92\%, 84\%, and 41\% for VGG11s on CIFAR-10, compared to the respective baselines (FedPer, FedBuff, FedAsync, FedAvg+Topk).


\begin{table}[!t]\scriptsize
\setlength{\belowcaptionskip}{-3mm}
\centering
\caption{Comparison of elapsed time and communication consumption for reaching the target accuracy in Non-IID setting. }
\label{table:niid}
\scalebox{0.8}{
\begin{tabular}{|c|c|c|c|c|}
\hline

\multirow{2}{*}{\makecell[c]{Model@Dataset}} & \multirow{2}{*}{\makecell[c]{Target Accuracy(\%)}} & \multirow{2}{*}{\makecell[c]{Method}} & \multirow{2}{*}{\makecell[c]{Elapsed Time(s)}} & \multirow{2}{*}{\makecell[c]{Communication Consumption(GB)}} \\ & & & &  \\

\hline
\hline

\multirow{15}{*}{\makecell[c]{CNN@FMNIST}} 
& \multirow{5}{*}{$65$} & FedLuck & $\mathbf{41.92}$ & $\mathbf{0.02}$\\
\cline{3-5}
& & FedPer & $64.44$ & $0.05$ \\
\cline{3-5}
& & FedBuff & $57.32$ & $0.06$ \\
\cline{3-5}
& & FedAsync & $66.66$ & $0.07$ \\
\cline{3-5}
& & FedAvg+Topk & $66.65$ & $\mathbf{0.02}$ \\
\cline{2-5}
& \multirow{5}{*}{$75$} & FedLuck & $\mathbf{77.92}$ & $\mathbf{0.04}$\\
\cline{3-5}
& & FedPer & $211.43$ & $0.12$ \\
\cline{3-5}
& & FedBuff & $141.17$ & $0.13$ \\
\cline{3-5}
& & FedAsync & $227.64$ & $0.24$ \\
\cline{3-5}
& & FedAvg+Topk & $199.95$ & $\mathbf{0.04}$ \\
\cline{2-5}
& \multirow{5}{*}{$85$} & FedLuck & $\mathbf{581.92}$ & $0.32$\\
\cline{3-5}
& & FedPer & $644.94$ & $0.36$ \\
\cline{3-5}
& & FedBuff & $614.88$ & $0.53$ \\
\cline{3-5}
& & FedAsync & $841.53$ & $0.87$ \\
\cline{3-5}
& & FedAvg+Topk & $1066.39$ & $\mathbf{0.21}$ \\
\hline

\multirow{15}{*}{\makecell[c]{VGG11s@CIFAR-10}} 
& \multirow{5}{*}{$50$} & FedLuck & $\mathbf{837.00}$ & $\mathbf{0.31}$\\
\cline{3-5}
& & FedPer & $1257.26$ & $0.76$ \\
\cline{3-5}
& & FedBuff & $3694.49$ & $5.13$ \\
\cline{3-5}
& & FedAsync & $2134.37$ & $2.81$ \\
\cline{3-5}
& & FedAvg+Topk & $2677.70$ & $0.65$ \\
\cline{2-5}
& \multirow{5}{*}{$60$} & FedLuck & $\mathbf{1484.99}$ & $\mathbf{0.55}$\\
\cline{3-5}
& & FedPer & $1959.25$ & $1.18$ \\
\cline{3-5}
& & FedBuff & $5979.43$ & $8.28$ \\
\cline{3-5}
& & FedAsync & $3163.80$ & $4.16$ \\
\cline{3-5}
& & FedAvg+Topk & $4471.76$ & $1.08$ \\
\cline{2-5}
& \multirow{5}{*}{$65$} & FedLuck & $\mathbf{1965.00}$ & $\mathbf{0.73}$\\
\cline{3-5}
& & FedPer & $2760.26$ & $1.66$ \\
\cline{3-5}
& & FedBuff & $8635.33$ & $11.94$ \\
\cline{3-5}
& & FedAsync & $4565.33$ & $6.00$ \\
\cline{3-5}
& & FedAvg+Topk & $5944.50$ & $1.44$ \\
\hline

\multirow{15}{*}{\makecell[c]{LSTM@SC}} 
& \multirow{5}{*}{$50$} & FedLuck & $\mathbf{1438.16}$ & $0.54$\\
\cline{3-5}
& & FedPer & $1754.43$ & $\mathbf{0.36}$ \\
\cline{3-5}
& & FedBuff & $1586.80$ & $1.48$ \\
\cline{3-5}
& & FedAsync & $2007.81$ & $2.39$ \\
\cline{3-5}
& & FedAvg+Topk & $2813.85$ & $0.63$ \\
\cline{2-5}
& \multirow{5}{*}{$60$} & FedLuck & $\mathbf{2436.00}$ & $0.91$\\
\cline{3-5}
& & FedPer & $3619.29$ & $\mathbf{0.73}$ \\
\cline{3-5}
& & FedBuff & $3176.19$ & $2.95$ \\
\cline{3-5}
& & FedAsync & $3124.58$ & $3.72$ \\
\cline{3-5}
& & FedAvg+Topk & $4592.93$ & $1.02$ \\
\cline{2-5}
& \multirow{5}{*}{$63$} & FedLuck & $\mathbf{2705.98}$ & $1.01$ \\
\cline{3-5}
& & FedPer & $4062.71$ & $\mathbf{0.82}$ \\
\cline{3-5}
& & FedBuff & $4418.68$ & $4.11$ \\
\cline{3-5}
& & FedAsync & $3604.80$ & $4.30$ \\
\cline{3-5}
& & FedAvg+Topk & $5409.86$ & $1.20$ \\
\hline
\end{tabular}
}

\vspace{-1.0em}
\end{table}

\textbf{The impact of Non-IID data.} In the experiments conducted above, we utilized only independent and identically distributed (IID) data. However, in federated learning, dealing with not identically and independently distributed (Non-IID) data is a critical concern that can decelerate convergence speed.  We further conducted tests on the impact of Non-IID data, in line with Assumption 2 (Eq. \ref{ass:noniid}). We generated Non-IID data using the Dirichlet distribution, randomly assigning it to each local device to create Non-IID data distributions across devices, with a concentration degree $\alpha$ of 1.0 \cite{direc}, reflecting real-world scenarios. Tab. 1 shows the results of our experiments with non-IID data. FedLuck continues to outpace the baselines in both convergence speed and communication consumption. For instance, FedLuck only uses 1965.00s to achieve the 65\% target testing accuracy on VGG11s training on CIFAR-10, while FedPer, FedBuff, FedAsync and FedAvg+Topk use 2760.26s, 8635.33s, 4565.33s, 5944.50s, respectively. FedLuck not only converged to 65\% target testing accuracy in the shortest possible time but also required less communication consumption. 

\begin{table}[!t]\footnotesize
    \centering
    \caption{Comparison of the top-1 test accuracy between FedLuck and methods that optimize only one factor over 800s for CNN, 6000s for VGG11s and LSTM. }
    \scalebox{1.05}{
    \begin{tabular}{cccc}
        \hline
         Top-1 Accuracy & \makecell[c]{CNN@FMNIST} & \makecell[c]{VGG11s@CIFAR-10} & \makecell[c]{LSTM@SC} \\
        \hline
         FedLuck & $\mathbf{86.9 \pm 0.4}\%$ & $\mathbf{78.9 \pm 0.2}\%$ & $\mathbf{71.0 \pm 0.4} \%$ \\
         Opt.~CR & $83.2 \pm 0.5\%$ & $67.6 \pm 0.4\%$ & $70.8 \pm 0.5 \%$ \\
         Opt.~LF & $81.5 \pm 0.4\%$ & $77.0 \pm 0.3\%$ & $69.8 \pm 0.3 \%$ \\
         \hline
    \end{tabular}
    }
    
    \vspace{-2.4em}
    
    \label{table:joint}
\end{table}

\textbf{The necessity of joint optimization.} To explore the essential "joint" characteristic of our FedLuck method, we conducted experiments and compared it with two separate baselines. In the first baseline Opt. CR, we adopted a fixed gradient compression rate and optimized only the local updating frequency. Conversely, in the second baseline Opt. LF, we kept the local updating frequency static and focused solely on optimizing the gradient compression rate. According to Tab. 2, FedLuck excels in all three tasks, demonstrating that the joint optimization of both local updating frequency and compression rate is beneficial in creating a more efficient AFL system.
\section{Conclusion}\label{conclusion}

In this paper, we tackle the inherent challenges of AFL arising from device heterogeneity and low-bandwidth communication environments, focusing on the issue of model staleness. We propose FedLuck, a novel AFL framework to achieve enhanced training efficiency by jointly and adaptively deciding the local updating frequency and compression rate. Through a thorough exploration of these factors, we derive a convergence upper bound directly linked to local update frequency and gradient compression rate, which allows us to carefully adjust these parameters to optimize the process. FedLuck demonstrates impressive results, significantly reducing communication consumption by 56\% and training time by approximately 55\%, making it a competitive alternative compared to existing baselines in scenarios with heterogeneous and low-bandwidth conditions. To our best knowledge, we are the first to jointly optimize AFL in this manner, thereby contributing to more efficient and effective federated learning in real-world settings. The insights and contributions presented in our paper pave the way for future research to refine and extend the approach, addressing other challenges in AFL and advancing the state-of-the-art in the field.

\section*{ACKNOWLEDGMENT}
This work is supported in part by the National Key Research and Development Project of China (Grant No. 2023YFF0905502), Shenzhen Science and Technology Program (Grant No. JCYJ20220818101014030).
\footnotesize
\bibliographystyle{splncs04}
\bibliography{reference}

\endgroup
\end{document}